\input{psfig.sty}

\documentstyle[11pt,aaspp4, flushrt]{article}  

\begin{document}
 
\title{The Discovery of a State Dependent Hard Tail in the X-ray Spectrum 
of the Luminous Z-source GX~17+2}

\author{T. Di Salvo\altaffilmark{1}, L. Stella\altaffilmark{2,3},
N. R. Robba\altaffilmark{1}, M. van der Klis\altaffilmark{4},
L. Burderi\altaffilmark{2}, 
G.L. Israel\altaffilmark{2,3}, J. Homan\altaffilmark{4},
S. Campana\altaffilmark{3,5}, F. Frontera\altaffilmark{6}, 
A.N. Parmar\altaffilmark{7}}
\altaffiltext{1}{Dipartimento di Scienze Fisiche ed Astronomiche, 
Universit\`a di Palermo, Via Archirafi 36, 90123 Palermo, Italy;
disalvo@gifco.fisica.unipa.it.}
\altaffiltext{2}{Osservatorio Astronomico di Roma, Via Frascati 33, 
00040 Monteporzio Catone (Roma), Italy; stella@ulysses.mporzio.astro.it.}
\altaffiltext{3}{Affiliated with the International Center for Relativistic 
Astrophysics.}
\altaffiltext{4}{Astronomical Institute "Anton Pannekoek," University of 
Amsterdam and Center for High-Energy Astrophysics,
Kruislaan 403, NL 1098 SJ Amsterdam, the Netherlands.}
\altaffiltext{5}{Osservatorio Astronomico di Brera, Via Bianchi 46, I-23807 
Merate, Italy.}
\altaffiltext{6}{Istituto Tecnologie e Studio Radiazioni Extraterrestri, 
CNR, Via Gobetti 101, 40129, Bologna, Italy.}
\altaffiltext{7}{Astrophysics Division, Space Science Department of ESA, 
ESTEC, P.O. Box 299, 2200 AG Noordwijk, Netherlands.}
  

\begin{abstract}
We report results of a BeppoSAX (0.1--200~keV) observation of 
the Z-type low mass X-ray binary GX~17+2.
The source was on the so-called Horizontal and Normal branches.
Energy spectra were selected based on the source position in the 
X-ray hardness-intensity diagram. 
The continuum could be fairly well 
described by the sum of a $\sim 0.6$~keV blackbody, 
contributing $\sim 10\%$ of the observed $0.1-200$~keV flux, and a 
Comptonized component, resulting from upscattering of $\sim 1$~keV seed 
photons by an electron cloud with temperature 
of $\sim 3$~keV and optical depth of $\sim 10$.
Iron K-line and edge were also present at energies
$\sim 6.7$ and $\sim 8.5$~keV, respectively.
In the spectra of the Horizontal branch a hard tail was clearly 
detected at energies above $\sim 30$~keV.  
It could be fit by a power law of photon index $\sim 2.7$, 
contributing $\sim 8\%$ of the source flux.
This component gradually faded as the source moved 
towards the Normal branch, where it was no longer detectable.
We discuss the possible origin of this component and the similarities with 
the spectra of Atoll sources and black hole X-ray binaries.
\end{abstract}


\section{Introduction}

Low Mass X-ray Binaries (hereafter LMXBs) hosting an old accreting neutron 
star (NS) display X-ray spectra that are still relatively poorly understood. 
The modern classification of LMXBs relies upon the 
branching displayed by individual 
sources in the X-ray color-color diagram (Hasinger \& van der Klis 1989),   
and comprises a Z-class
(source luminosities close to the Eddington luminosity, $L_{\rm edd}$) and 
an Atoll-class (usually luminosities of $\sim 0.01-0.1\ L_{\rm edd}$).
Considerable evidence has been found that the mass accretion rate 
of individual Z-sources increases from the top left to the 
bottom right of the Z-pattern ({\it e.g.} Hasinger et al. 1990), 
i.e. along the so called horizontal, normal and flaring branches 
(hereafter HB, NB and FB, respectively). 
While the spectra of Atoll sources often extend up 
to energies of $\geq 100$~keV and present interesting analogies 
with the hard X-ray component of black hole candidate (BHC) binaries,
the X-ray spectra of Z sources are comparatively soft 
and dominated by thermal-like components with exponential cutoff 
at energies $\le 10$~keV. 
Attempts at decomposing the X-ray spectra of Z sources have employed 
two (or more) components, the 
origin of which is still debated ({\it e.g.} Mitsuda et al. 
1984; White et al. 1986; 1988; Psaltis, Lamb, \& Miller 1995). A tracking 
of the spectral variations along the Z pattern was attempted 
(see {\it e.g.} Hasinger et al. 1990; Hoshi \& Mitsuda 1991;
Schulz \& Wijers 1993; Asai et al. 1994, and references therein); the results 
were inconclusive, since in different decompositions the 
spectral variations can be ascribed to different components.
Moreover it is not clear whether any of the proposed components 
evolve with continuity from Z to Atoll sources.
The timing properties of LMXBs, on the contrary, possess 
a clear continuity along the different branches of each source and also 
a remarkable similarity across Z and Atoll sources (e.g.  Wijnands \& 
van der Klis 1999; Psaltis, Belloni, \& van der Klis 1999).

We report here the results of an extensive study with BeppoSAX 
of the Z source GX~17+2.
The source usually describes a complete Z-pattern in the color-color 
diagram in a relatively short time (3--4 days). 
The parameters of the noise components as well as the frequency of the QPOs
are well correlated with the position in the color-color diagram 
({\it e.g.} Langmeier et al. 1990; Kuulkers et al. 1997; Wijnands et al. 1997).
The source occasionally shows X-ray bursts which are often long 
and involve a moderate increase of the 
X-ray flux ($\sim 50$\%; Kahn \& Grindlay 1984;  
Tawara et al. 1984); Sztajno et al. (1986) interpret them as type-I X-ray 
bursts. 
Studies of the source X-ray spectrum were carried out by White, Stella, 
\& Parmar (1988), Langmeier et al. (1990), Schulz \& Wijers (1993), 
based on EXOSAT spectra, and Hoshi \& Asaoka (1993) based on Ginga spectra.
In all cases the spectrum was found to steepen rapidly above energies 
of a few tens of keV.

\section{Observations and Analysis}

A pointed observation of GX~17+2 was carried out 
between 1999 October 4 04:52 UT and October 9 01:10 UT, with the 
Narrow Field Instruments, NFIs, on board BeppoSAX 
(Boella et al. 1997).  These consist of four co-aligned instruments covering
the 0.1--200~keV energy range: LECS (0.1--10~keV), two MECS (1.3--10~keV),
HPGSPC (7--60~keV), and PDS (13--200~keV).
The effective exposure was 56~ks in the LECS, 193~ks in the
two MECS, 160~ks in the HPGSPC, and 78~ks in the PDS.
In the LECS and MECS images data were extracted from 
circular regions centered on GX~17+2, with radii of $8'$ and $4'$ respectively.
Data extracted from the same detector regions during blank field 
observations were used for background subtraction.
Background subtraction was obtained in the PDS from data accumulated during 
the off-source intervals, and in the HPGSPC by using spectra accumulated from
Dark Earth data.
All the available evidence indicates that there were no significant 
contaminating sources in the field of view of the HPGSPC and PDS:
the off-source PDS count rates are in the expected range;
the HPGSPC and PDS spectra align well with each other and with the MECS
spectra; finally, evidence for a hard X-ray component in GX~17+2 was found 
only when the source was in a specific state (see below). 

Four bursts occurred during our observation; these
were excluded from the subsequent analysis.
Figure 1 shows the Hardness-Intensity
Diagram, HID, where the Hard Color, HC, is the ratio of the 7--10.5
keV to the 4.5--7~keV MECS count rates and the intensity is the
4.5-10.5~keV count rate. The HB and NB are clearly seen.
A comparison with a longer RXTE observation, simultaneous to the BeppoSAX 
observation, shows that the BeppoSAX HID covers the entire HB and NB
(Homan et al. 2001). 
In order to investigate the source spectrum in different regions of 
the HID, we accumulated the energy spectra of each NFI over the seven different
HC intervals defined by the following boundaries: 0.26, 
0.288, 0.3, 0.31, 0.315, 0.32, 0.335, 0.37. These spectra are indicated
with A, B, ..., G in order of decreasing HC. 
A systematic error of 1\% was added to the data of each spectrum. 
As customary, in the spectral fitting procedure we allowed for different
normalizations in the LECS,  HPGSPC and PDS spectra relative to the MECS
spectra, and checked  {\it a posteriori} that derived values are in the
standard range for each instrument
(see http://www.sdc.asi.it/software).

In line with previous studies (Mitsuda et al. 1984; White et al. 1988), 
we tried several two-component models to fit the X-ray continuum of GX~17+2.
These were a blackbody or a disk multicolor blackbody ({\tt diskbb}, 
Mitsuda et al. 1984) for the soft component, together with a blackbody or 
models of thermal Comptonization, such as a power law with cutoff,  
{\tt compst} (based on the solution of the Kompaneets equation given by 
Sunyaev \& Titarchuk 1980), 
or {\tt comptt} (in which relativistic effects are included
and the dependence of the scattering opacity on seed photon energy, a 
free parameter of the model, is taken into account, Titarchuk 1994), 
for the harder component.  
In all cases the addition of a Fe K$\alpha$ line at $\sim 6.7$~keV 
proved necessary (probability of chance improvement $\sim 10^{-7}-10^{-8}$), 
while an edge at $\sim 8.5$~keV improved the fits 
at $\sim 98-99\%$ confidence level.
The best fit to the continuum of GX~17+2 was obtained using
a blackbody plus the {\tt comptt} Comptonization model
(reduced chisquare, averaged on the intervals, of $\chi^2_r \sim 1.26$ vs. 
$\chi^2_r \sim 1.31-1.51$ for the other models).
The blackbody had a temperature of 
$kT_{\rm BB}\sim 0.6$~keV and contributed $\sim 10$\,\% of the observed 
0.1-200~keV flux. 
The blackbody equivalent radius was $R_{\rm BB}\sim 37$~km  
(using a distance of 7.5 kpc, Penninx et al. 1988). 
The temperature of the seed photons for the Comptonized component was 
$kT_{\rm W}\sim 1$~keV.  Following In 't Zand et al. (1999), we derived 
the effective Wien radius of the seed photons; this was 
$R_{\rm W} = 12-17$~km. 
The temperature and optical depth of the (spherical) Comptonizing 
region were $k T_{\rm e} \sim 3$~keV and $\tau \sim 10-12$, respectively.  

A marked excess of counts above $\sim 30$~keV was clearly apparent in the 
spectrum from the upper HB (spectrum A, $HC > 0.335$).  This excess
could not be fit by any of the 
two-component continuum models that we tried.  
Therefore we added an extra component to the model above. 
With the addition of a power law the  
$\chi^2$ decreased from 339 (for 239 d.o.f) to 254 (237 d.o.f.) in the 
case of spectrum A. An
F-test indicates that the probability of chance improvement is negligible
($\sim 10^{-13}$). This additional component was detected up to energies 
of $\sim 100$~keV, had a best fit photon index $\sim 2.7$ and  
contributed  $\sim 8\%$ of the 0.1-200~keV source flux. 
Other models ({\it e.g.} a thermal 
bremsstrahlung with  $kT \sim 30$~keV) gave comparably good results.  
The addition of a power law component, with photon index 2.7, 
also improved
the fit of spectra B, C and D (from the lower HB, $0.31 < HC < 0.33$). 
We note that the source went back several times to the same regions of 
the HB, each time with the same effect on the hard tail.
Results from these three-component fits to 
the BeppoSAX spectra of GX~17+2 are given in Table~1.

On the contrary, the spectra from the NB (spectra E, F and G) did not 
show any evidence for a high energy excess.
To study the flux variations of the high energy component 
across the HB and NB we fixed the  photon index to 2.7 ({\it i.e.} the 
best fit value for spectrum A) and derived upper limits on the normalization.  
It is apparent from Table 1 that the flux in the high energy power law 
must have decreased by at least a factor of $\sim 20$, from the upper HB 
(spectrum A) to the lower NB (spectrum G). See also Figure 2, which 
shows spectra A and G 
and the residuals with respect to the corresponding best fits.

\section{Discussion}

Our observation of GX~17+2, the first to extensively study the X-ray spectrum 
of a Z source with sensitive instrumentation covering a broad energy range, 
led to 
the discovery of a high energy 
component extending up to $\sim 100$~keV, at least. 
This component was detected in the HB of the source,
where it contributed up to $\sim 8$\% of the 0.1--200~keV source flux. 
On the contrary the NB spectrum did not show any evidence for a high energy 
excess (upper limit of $\sim 0.4$\% in the lower NB).
Our modelling of the X-ray continuum up to energies of $\sim 20-30$~keV 
comprises a blackbody, contributing $\sim 10$\% of the 
0.1--200~keV flux, and a Comptonized spectrum, resulting from the 
upscattering of seed photons emitted by a $\sim 1$~keV Wien spectrum. 
This continuum model might be regarded as a generalization of the 
so-called ``western" model, with several important differences 
(see White et al. 1986, 1988):
(a) the blackbody has a lower temperature
($\sim 0.6$ vs. $\sim 1-2$~keV) and contributes a lower fraction of the 
X-ray flux ($\sim 10$\% vs. $\sim 10-40$\%); the corresponding
effective radii are $\sim 40$~km, i.e. larger than 
the NS radius; 
(b) the seed photons of the Comptonized spectrum are emitted at 
X-ray energies (as opposed to UV energies), perhaps by the NS surface
or boundary layer, or by the magnetosphere of the NS for self-absorbed 
cyclotron emission (see Psaltis et al. 1995).
The spectral evolution across the HB to the lower NB can be ascribed 
mainly to a monotonic decrease of the electron temperature and optical depth
of the Comptonized component (see Table 1), which, within the spectral 
decomposition adopted here, are mainly responsible 
for the decrease of the HC from the HB to the NB. 
On the contrary, the soft part of the spectra remains nearly unchanged, 
in agreement with the fact that the HB and NB of GX~17+2
are almost vertical in the color-color diagram (soft 
color on the x-axis, see Wijnands et al. 1997).
A more detailed discussion of the new continuum model for the X-ray spectrum 
of GX~17+2 in relation to the properties of the X-ray bursts from the
source will be presented elsewhere. 
We note that the same continuum model 
has been used to describe the X-ray spectrum of Atoll sources 
({\it e.g.} Guainazzi et al. 1998; Barret et al. 2000).
A broad iron emission line is also present at $\sim 6.7$~keV, 
with a FWHM of 0.4--0.6~keV and an equivalent width of $\sim 40$ eV;
these values are similar to those reported by White et al. (1986). 
The detection of an absorption edge at $\sim 8.5$~keV is new. The 
energy of the line and absorption edge indicate a high ionization stage 
of Fe (Fe XXIII--XXV).

The main result of our study is the discovery of a hard 
component in the X-ray spectrum of GX~17+2. This component
was detected only in the HB, dominating the spectrum
between $\sim 40$ and $\sim 100$~keV. 
A $\Gamma \simeq 2.7$ power law model provided an adequate fit. 
This is the first time that a hard X-ray component is detected in the 
HB of a Z-source. Previously a power law-like hard excess was detected in the 
Ginga data of GX~5--1 (Asai et al. 1994). GX~5--1 was in the NB and FB during 
the {\it Ginga} observation, and the intensity of the power law component
decreased from the NB to the FB. 
A hard X-ray component was also detected in the BeppoSAX data of Cyg~X-2,
but the source state was not reported (Frontera et al. 1998). 
These results indicate that hard components are probably a common feature
of Z sources and that their intensity decreases from the HB to the NB and the 
FB, {\it i.e.} for increasing mass accretion rates. 
We note that this behavior is similar to that observed in LMXBs of the 
Atoll group and BHCs, where the relative importance of the hard X-ray 
component is often highest in the low-luminosity source states
(the island state of Atoll source and the low and intermediate states of BHCs, 
see {\it e.g.} Nowak 1995; Barret et al. 2000). 
Theoretical interpretations for this effect have already been 
discussed: for example at high accretion rates the corona might 
be efficiently cooled 
by increased Comptonization losses, or swept away by  
radiation pressure (Popham \& Sunyaev 2000).  

Di Matteo \& Psaltis (1999)
found an interesting correlation between the QPO frequency and  
the slope of the power-law energy spectra in BHCs.
If the QPO frequency is related to the size of the innermost 
(optically thick, geometrically thin) disk region, 
that changes in response to accretion rate variations (as indeed envisaged 
in a number of QPO models, see van der Klis 2000 for a review), then 
this correlation might simply reflect variations of the Comptonization 
parameter $y$ in response to variations of the innermost disk radius. 
We note that, owing to poor statistics, 
the BeppoSAX GX~17+2 data presented here do not 
provide independent evidence that the hard component 
steepens as the source moves from the upper HB and to the NB,
{\it i.e.} the path along which the QPO frequency increases. 
Yet the data are consistent with this behavior.
We note that there might exist a relationship between the 
presence of the hard X-ray component and QPOs. This is suggested by the 
fact that: 
(a) the hard X-ray component of GX~17+2 is most 
pronounced over the source state(s) in which both the kHz and HB
QPOs are detected and reach the highest rms amplitudes;  
(b) both these rms amplitudes and the contribution from the 
hard X-ray component to the total spectrum increase dramatically 
with energy (Wijnands et al. 1997).

The spectrum of GX~17+2 in the HB is similar to the spectra of BHCs
in some soft states (intermediate or very high states). These are 
dominated by soft emission with characteristic temperatures of 
$\sim 1-2$~keV, and display a hard power-law component of photon 
index $\sim 2-3$, extending to several hundreds keV and contributing
a few per cent of the total luminosity (see e.g. Grove et al. 1998).
At a purely phenomenological level, 
the presence of a similar high energy tail in an otherwise soft 
NS system indicates that this tail should not be considered a signature 
of BHC accretion.
The origin of this extended power-law tail in some soft states of BHCs 
is still debated.
A model that might explain the hard power-law component in the soft
state of both NS and BHC systems is the hybrid 
thermal/non-thermal corona (see e.g. Gierli\'nski et al. 1999), where
a fraction of the energy can be injected in form of electrons with 
a non-thermal velocity distribution.
On the other hand the bulk-motion Comptonization model (Ebisawa, Titarchuk, 
\& Chakrabarti 1996) is unlikely to work in luminous NS systems
where the radial bulk motion would be slowed down by the radiation emitted 
close to the NS surface.

An alternative possibility is that the hard X-ray component in GX~17+2
originates from the scattering off non-thermal electrons
of a jet, that is probably present in the source at low 
accretion rates. 
In fact GX~17+2 is associated with a variable radio source,
the flux of which decreases from the  
HB to the FB (Penninx et al. 1988).  This is a general characteristic of
Z sources, but not of Atoll sources
(Hjellming \& Han 1995; Fender \& Hendry 2000, and references therein).

\acknowledgments
We thank D. Psaltis for useful suggestions.
This work was partially 
supported by the Italian Space Agency (ASI) and the Ministero
della Ricerca Scientifica e Tecnologica (MURST).

\clearpage

\section*{TABLE}

\begin{table}[h!]
\tiny
\begin{center}
\caption{Results of the fit of GX~17+2 spectra in the energy band 
0.12--200~keV. 
The model consists of blackbody, 
{\tt comptt}, power law, a Gaussian emission line, and an absorption edge.
For each spectrum the corresponding range of HC is indicated.
Uncertainties are at the 90\% confidence level for a single parameter.
The power-law normalization is in units of ph keV$^{-1}$ cm$^{-2}$ s$^{-1}$ 
at 1~keV.
The total flux
is in the 0.1--200~keV energy range.
F-test is the probability of chance improvement when the power law is
included in the spectral model.
}

\vskip 0.5cm
\label{tab1}
\begin{tabular}{l|c|c|c|c|c|c|c} 
\tableline \tableline
 Spectrum  & A   &    B        &     C       &     D       &       E     &    F 
        
&      G       \\
 Hard Color  & (0.335--0.37) &(0.32--0.335)&(0.315--0.32)&(0.31--0.315)& (0.3--
0.31) & (0.288--0.3) & (0.26--0.288) \\
\tableline

$N_{\rm H}$ $\rm (\times 10^{22}\;cm^{-2})$
& $2.2 \pm 0.2$  & $2.12 \pm 0.08$ & $1.92 \pm 0.07$
& $1.97 \pm 0.06$ & $1.83 \pm 0.06$
& $1.80 \pm 0.05$ & $1.85 \pm 0.06$ \\

$k T_{\rm BB}$ (keV)
& $0.63 \pm 0.06$ & $0.52 \pm 0.04$ & $0.58 \pm 0.04$
& $0.57 \pm 0.04$ & $0.54 \pm 0.04$
& $0.56 \pm 0.04$ & $0.56 \pm 0.04$ \\

$R_{\rm BB}$ (km)
& $35 \pm 7$        & $47 \pm 8$      & $38 \pm 6$ 
& $40 \pm 6$        & $43 \pm 7$ 
& $41 \pm 6$        & $41 \pm 6$ \\

$k T_{\rm W}$ (keV)
& $1.07 \pm 0.09$ & $0.95 \pm 0.05$ & $1.01 \pm 0.05$
& $1.01 \pm 0.05$ & $1.01 \pm 0.03$
& $1.03 \pm 0.04$ & $1.03 \pm 0.04$ \\

$k T_{\rm e}$ (keV) 
& $3.28 \pm 0.06$ & $3.21 \pm 0.05$ & $3.25 \pm 0.05$
& $3.15 \pm 0.04$ & $3.08 \pm 0.04$
& $3.13 \pm 0.04$ & $3.04 \pm 0.04$ \\

$\tau$
& $11.6 \pm 0.4$ & $11.0 \pm 0.3$ & $10.6 \pm 0.2$
& $10.6 \pm 0.2$ & $9.8 \pm 0.2$
& $9.9 \pm 0.2$ & $9.5 \pm 0.2$ \\

R$_{\rm W}$ (km)
& $12 \pm 2$  &  $17 \pm 2$  &  $15 \pm 1$
& $15 \pm 1$ & $16 \pm 1$
& $15 \pm 1$ & $15 \pm 1$ \\

Photon Index 
& $2.7^{+0.2}_{-0.3}$ & 2.7 (fixed) & 2.7 (fixed)  
& 2.7 (fixed) & 2.7 (fixed) 
&  2.7 (fixed) &  2.7 (fixed) \\

Power-law N
& $1.3 \pm 0.7$ & $0.6 \pm 0.2$ & $0.4 \pm 0.2$  
& $0.4 \pm 0.1$ & $< 0.27$
& $< 0.25$ &  $< 4.3 \times 10^{-2}$ \\

$E_{\rm Fe}$ (keV)
& $6.74 \pm 0.09$ & $6.78 \pm 0.09$ & $6.7 \pm 0.1$
& $6.7 \pm 0.1$ & $6.7 \pm 0.1$
& $6.7 \pm 0.1$ & $6.74 \pm 0.08$ \\

FWHM (keV)
& $0.6 \pm 0.3$ & $0.4 \pm 0.3$ & $0.5 \pm 0.3$ 
& $0.5 \pm 0.3$ & $0.5 \pm 0.4$
& $0.5 \pm 0.3$ & $0.5 \pm 0.4$ \\

I$_{\rm Fe}$ (ph cm$^{-2}$ s$^{-1}$)
& $(6 \pm 2) \times 10^{-3}$ & $(5 \pm 1) \times 10^{-3}$ & 
$(5 \pm 1) \times 10^{-3}$ & $(4 \pm 1) \times 10^{-3}$ 
& $(4 \pm 1) \times 10^{-3}$ & $(4 \pm 1) \times 10^{-3}$ 
& $(5 \pm 1) \times 10^{-3}$ \\

Fe Eq. W. (eV)
&  41 & 30 & 31 & 30 & 30 & 29 & 39  \\

$E_{\rm edge}$ (keV)
& $8.7 \pm 0.3$ & $8.8 \pm 0.3$ & $8.5 \pm 0.3$
& $8.6 \pm 0.3$ & $8.6 \pm 0.3$
& $8.5 \pm 0.3$ & $8.5 \pm 0.2$    \\

$\tau_{\rm edge}$ ($ \times 10^{-2}$)
& $3 \pm 1$ & $3 \pm 1$ & $3 \pm 1$
& $3 \pm 1$ & $2 \pm 1$
& $2 \pm 1$ & $5 \pm 1$  \\

$F_{\rm tot}$ (ergs cm$^{-2}$ s$^{-1}$)
& $1.84 \times 10^{-8}$   & $1.82 \times 10^{-8}$ & $1.81 \times 10^{-8}$ & 
$1.81 \times 10^{-8}$ & $1.69 \times 10^{-8}$ & $1.71 \times 10^{-8}$ &  
$1.52 \times 10^{-8}$    \\

$\chi^2_{\rm red}$ (d.o.f.)
& 1.07 (237) & 1.21 (238) & 1.04 (237)
& 1.11 (237) & 1.27 (236)
& 1.26 (237) & 1.22 (236) \\

F-test
& $1.2 \times 10^{-13}$  & $1.1 \times 10^{-4}$ & $4.9 \times 10^{-2}$ 
& $2.1 \times 10^{-4}$ & 0.14
& 0.30 &  0.98   \\

\tableline
\end{tabular}
\end{center}
\end{table} 

\clearpage
 
\begin{figure}[h]
\centerline
{\psfig
{figure=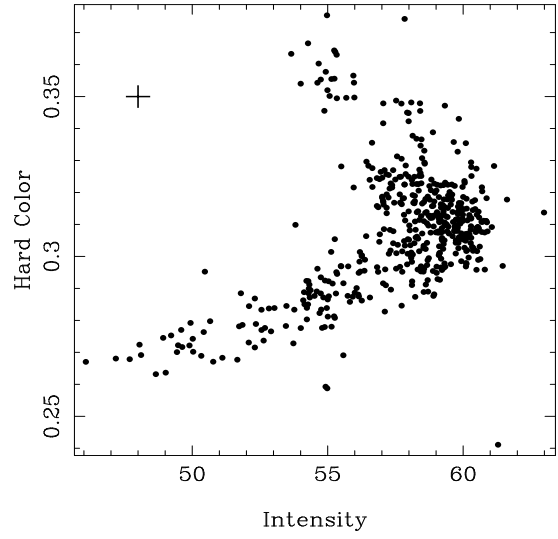,height=10.0cm,width=8.0cm}}
\caption{Hardness-Intensity Diagram of GX~17+2.
HC is the ratio of the counts in the 7--10.5~keV and 
4.5--7~keV energy bands. Each point corresponds to a 400 s integration.
A typical error bar is shown at the top left corner.}
\label{fig1}
\end{figure}

\begin{figure}[h]
\centerline
{\psfig
{figure=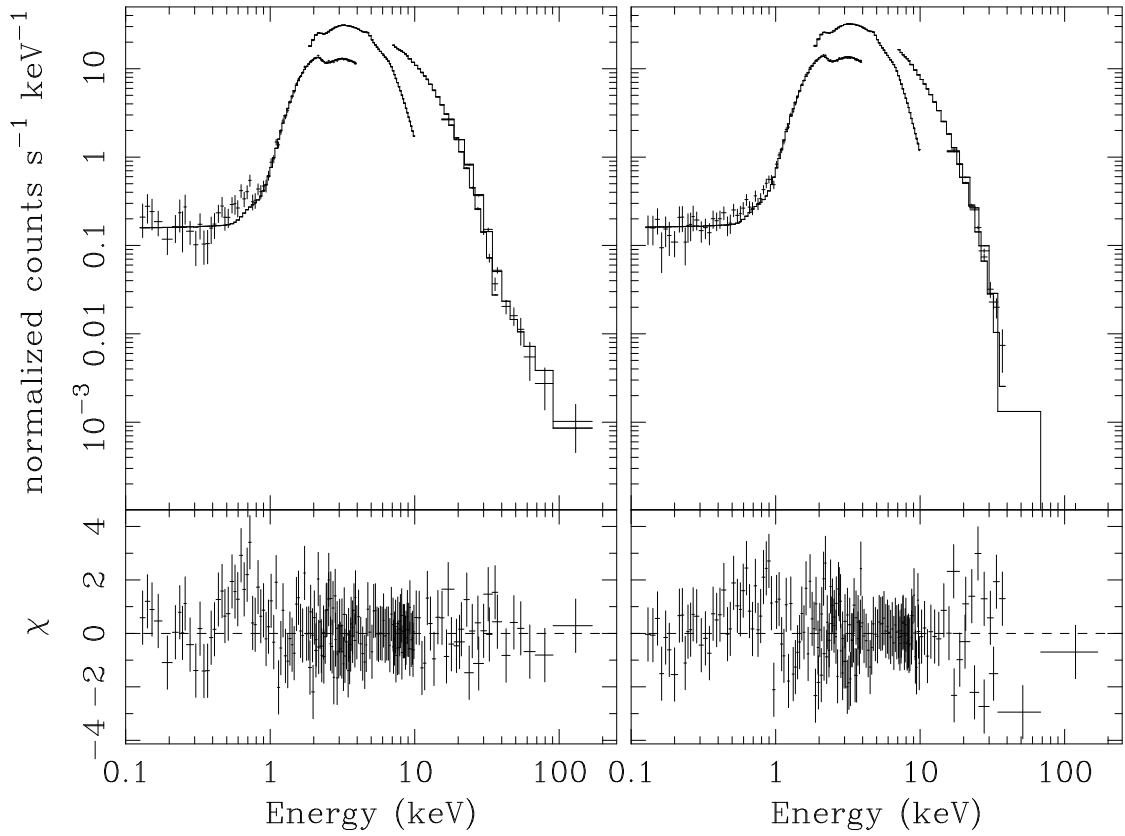,height=12.0cm,width=16.0cm,angle=270}}
\caption{Spectra A (upper HB, left panels) and G (lower NB,
right panels) and the corresponding best fit models (upper panels),
and residuals in unit of $\sigma$ (lower panels). }
\label{fig2}
\end{figure}

\end{document}